\documentclass[conference]{IEEEtran}
\usepackage{graphicx}
\usepackage[cmex10]{amsmath}
\usepackage{algorithmic}
\usepackage{array}
\usepackage[tight,footnotesize]{subfigure}
\usepackage{fixltx2e}
\usepackage{url}

\begin{document}

\title{Experimental study of continuous variable quantum key distribution}

\author{\IEEEauthorblockN{N. Benletaief, H. Rezig *Members IEEE, A. Bouallegue *Members IEEE}
\IEEEauthorblockA{Communication System laboratory Sys'Com\\
 National Engineering School of Tunis\\
 BP 37, 1002 Tunis  Belv\'ed\`ere, Tunisia\\
 benletaief.nedra@gmail.com, houria.rezig@enit.rnu.tn, ammar.bouallegue@enit.rnu.tn}

}

\maketitle

\begin{abstract}
It has been proven in the literature that the main technological factors limiting the communication rates of quantum cryptography systems by single photon are mainly related to the choice of the encoding method. In fact, the efficiency of the used sources  is very limited, at best of the order of a few percent for the single photon sources and the photon counters can not be operated beyond a certain speed and with a low order of detection efficiency. In order to overcome partially these drawbacks, it is advantageous to use continuous quantum states as an alternative to standard encodings based on quantum qubits. In this context, we propose a new reconciliation method based on Turbo codes. Our theoretical model assumptions are supported by experimental results.  Indeed, our method leads to a significant improvement of the protocol security and a large decrease of the QBER. The gain is obtained with a reasonable complexity increase. Also, the novelty of our work is that it tested the reconciliation method on a real photonic system under VPItransmissionMaker.
\end{abstract}


\IEEEpeerreviewmaketitle

\section{Introduction}
\begin{figure}[h]
\centering
\includegraphics[width=8cm,height=5cm]{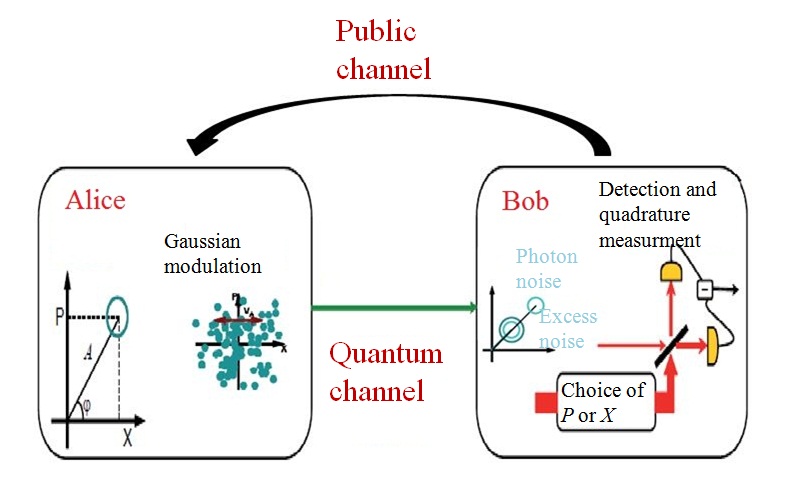}
\caption{QKD systems using continuous variable quantum state modulation.\label{prot}}
\end{figure}

Key distribution is an essential step for many classical cryptographic systems. It Allows two remote correspondents traditionally named Alice and Bob to establish a common secret key in order to encrypt their communications.
\\Among all proposed methods, only QKD provides  the  protocol with potentially unconditional security. Indeed, the major profit of QKD is the design of a physical solution to the problem of key distribution. Among the protocols of QKD, we Have the BB84 protocol \cite{Bennett}. It is Certainly the most famous and the most experienced of quantum cryptography. The simple proof of its security with respect to arbitrary eavesdropping strategies is given by Mayers \cite{prof3}  and was demonstrated later by Shor and Preskill \cite{prof4}. The protocol requires quantum transmission, construction of two correlated strings called sifted key, public comparison of  data in order to estimate the errors' parameters, reconciliation and finally amplification of confidentiality.
\\In the presence of an eavesdropper, environmental constraints acting on the transmission channel and the imperfections of experimental systems, the exchanged sequences are different. It clearly appears that reconciliation should be performed in such a situation in order to remove the errors as for the legitimate partners. Reconciliation impact both the safety and robustness of the protocol.
\\Although most of QKD systems use discrete modulation of quantum states, recent protocols use continuous variable quantum states modulation (see Figure \ref{prot}). Indeed, since 1999, more precisely in the following Ralph study \cite{ralph}, many protocols exploiting the quadratures electromagnetic field ($X$ and $P$) were analyzed.
\\Our work aims at finding a solution to the problem of reconciliation of continuous variable QKD protocol.  In section 2, we present an overview of previous related works. In Section 3, we describe our proposed solution to the problem of reconciliation. Section 4 will provide a discussion of our proposed method in terms of security, correcting ability and complexity. Also, we present selected experimental results of VPItransmissionMaker.

\section{Related works}
The efficiency of the reconciliation algorithms fixes in practice the achievable communication rates and distances of quantum cryptography systems for continuous variables. The design of an efficient and specific reconciliation algorithm for continuous variables is therefore absolutely necessary to allow these new systems to compete with their discrete analog. Research in this subject is abundant with two competing solutions that we present in the following.
\subsection{Sliced Error Correction (​​SEC)}
The method of reconciliation ​​SEC, proposed by GV Assche et \emph{al.} in \cite{LDPC} and \cite{continu1}, has achieved the first quantum key distribution system using continuous variables \cite{continu2}. It makes use of interactive correcting errors codes asymptotically efficient. The principle of this method is to convert Alice and Bob's symbols into bits and then to apply standard Binary Correction Protocol (BCP) for correcting errors and to take advantage of all available information to minimize the number of exchanged messages.
In theory, this optimal procedure shows its limits in terms of efficiency when the codes are of finite size. Also, we know that BCP is optimized for a Binary symmetric channel (BSC) and such we can lose the efficiency of the protocol if we are in the presence of other type of channel.  Then, it does not consider transmission over long distances($<10$ kilometers). It must be emphasized that reconciliation  by SEC only requires  exchange between the two parities. The algorithm therefore has a low computational complexity. Reconciliation by SEC is recognized efficient for an $SNR$ approximately 3 but its performance degrades in the case of low $SNR$ because it mainly breaks the symmetry of the gaussian problem.
\subsection{Reconciliation by LDPC Codes}
 In \cite{bib11}, authors investigated a new method of reconciliation inspired of coding techniques with LDPC codes. In order to achieve reconciliation, they used an Multilevel Coded Modulation (MLC) \cite{MLC}.
Although the MLC modulation is designed according to the rule of distance-balanced thereby, it gives an excellent asymptotic coding gain. In practice, the performance of these systems is severely degraded due to the high error rate for low levels. The work of the authors in \cite{bib11} has given results in terms of efficiency improving slightly those of the SEC with a great complexity. It is important to emphasize that this algorithm is valid regardless of the distribution of continuous variables, but there is no guarantee that it is possible to build efficient LDPC codes in all cases.
\\We can summarize what has been presented by: with less complexity, the reconciliation by SEC assures  almost the same performance.
\\This  study of the state of the art of reconciliation methods in the case of continuous variables, allows us to consider in the next section our own solution which is based on a completely new approach and is an application of an associated Turbo coding.
\section{Proposed method for continuous variable QDK reconciliation}
\begin{figure}[h]
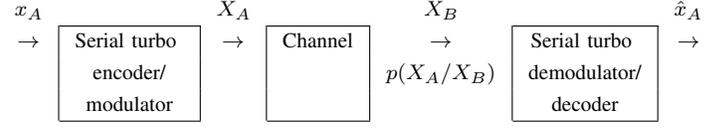

\centering
\begin{center}
{\footnotesize }%
\begin{tabular}{c|c|c|c|c|c|c}
\multicolumn{1}{c}{{\footnotesize $x_{A}$}} & \multicolumn{1}{c}{} & \multicolumn{1}{c}{{\footnotesize $X_{A}$}} & \multicolumn{1}{c}{} & \multicolumn{1}{c}{{\footnotesize $X_{B}$}} & \multicolumn{1}{c}{} & {\footnotesize $\hat{x}_{A}$}\tabularnewline
\cline{2-2} \cline{4-4} \cline{6-6}
{\footnotesize $\rightarrow$} & {\footnotesize Serial turbo } & {\footnotesize $\rightarrow$} & {\footnotesize Channel} & {\footnotesize $\rightarrow$} & {\footnotesize Serial turbo } & {\footnotesize $\rightarrow$}\tabularnewline
 & {\footnotesize encoder/} &  &  & {\footnotesize $p(X_{A}/X_{B})$} & {\footnotesize demodulator/} & \tabularnewline
 & {\footnotesize modulator} &  &  &  & {\footnotesize decoder} & \tabularnewline
\cline{2-2} \cline{4-4} \cline{6-6}
\multicolumn{1}{c}{} & \multicolumn{1}{c}{} & \multicolumn{1}{c}{} & \multicolumn{1}{c}{} & \multicolumn{1}{c}{} & \multicolumn{1}{c}{} & \tabularnewline
\end{tabular}
\par\end{center}
\caption{Reconciliation based on Turbo codes.\label{Fig:c0}}
\end{figure}

\begin{figure}[h]
\centering
\includegraphics[width=9cm,height=2.5cm]{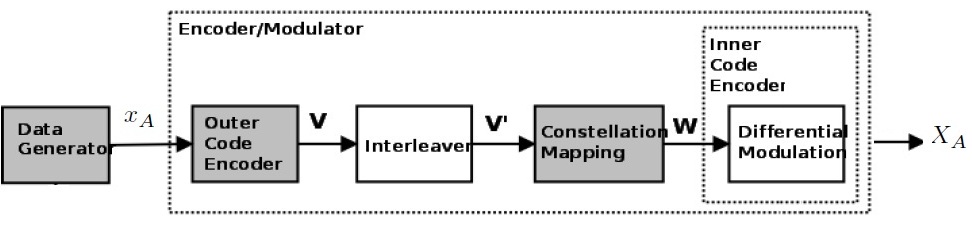}
\caption{Serial turbo encoder/modulator.\label{Fig:c1}}
\end{figure}

\begin{figure}[h]
\centering
\includegraphics[width=9cm,height=5cm]{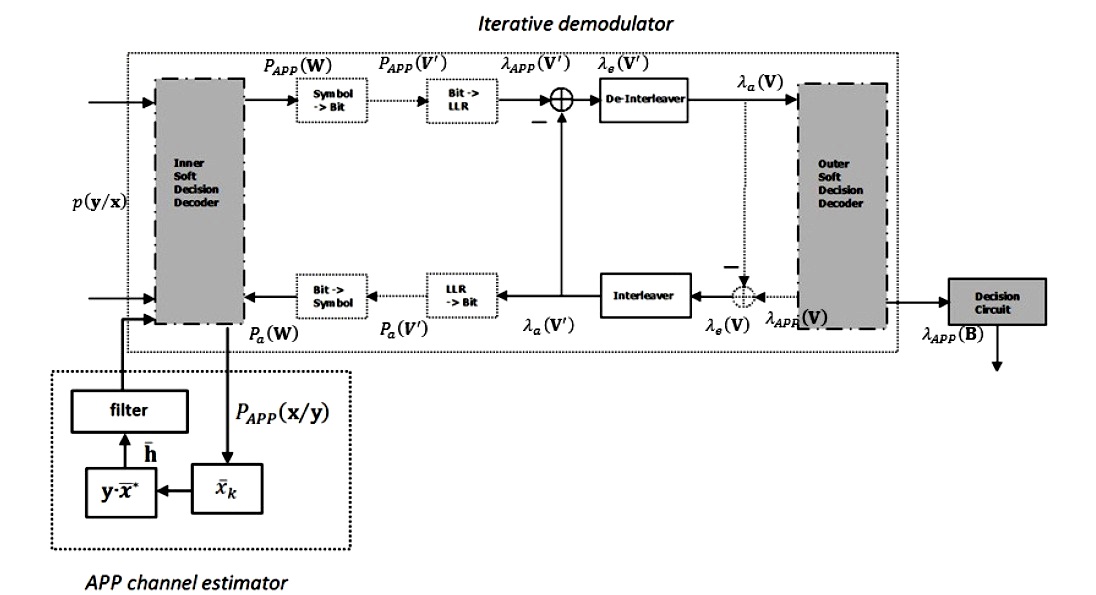}
\caption{Serial turbo demodulator/decoder \cite{dpsk}.\label{Fig:c2}}
\end{figure}

In this section, we will consider the special case of reconciliation of two continuous and gaussian quantum variables $X_{A}$ and $X_{B}$ ($X_{A}$ can be substituted by $P_{A}$, the other choice of quadrature, without loss of generality). $X_{A}$ is the sifted key at Alice's side and $X_{B}$ is the measured sifted key at Bob's side. We decided to adapt a given model in a whole other context \cite{dpsk}.
\\We then consider the model of Figure \ref{Fig:c0}.We generate random bits and then we encode them with a Turbo code. The new approach to handle continuous variables is to modulate them by differential phase shift keying (DPSK).
\\The coding/modulation block of the Figure \ref{Fig:c1} is composed of an outer encoder of a rate $\frac{2}{3}$, an interleaver, a PSK modulation and a differential modulator. First, a sequence of binary information $x_{A}$ of length $k$ is encoded into a string $V$ of size $M = \frac{3k}{2}$. Then, the coded bits are interleaved by a random interleaver, which gives the string $V^{'}$. After, this interleaved string is modulated into a string of symbols $W$ of length $S=\frac{k}{2}$. Finally, the modulated string undergoes differential encoding. The symbol transmitted is then $X_{A_{i}}=W_{i}X_{A_{i-1}}$. The string $X_{A}$ is of length $R= S + 1$. Differential modulation acts as an inner coder and it is seen as a recursive and non-systematic encoder.
 \\On the other side of the channel, we decode them according to the layout of Figure \ref{Fig:c2}. Thus, both parties can share a common key without having to discretize the signal as in the reconciliation SEC.

In our case study the channel is gaussian, so we can make the assumption that
\begin{equation}
  X_{B}=X_{A}+\epsilon
\end{equation}
where $\epsilon\sim\mathcal{N}(0,1+G\zeta)$ is a gaussian noise with variance $1+G\zeta$.
 $\zeta$ is an excess noise and $G$ is the total transmission of the quantum channel.
  \\The string $X_{B}$ is the input of an internal decoder that uses the \emph{a posteriori} probability. This decoder estimates the original symbol of the string $W$ by calculating the marginal \emph{a posterior} probability of each modulated symbol $W_ {i}$ knowing the sequence $X_ {B}$. The complete decoding algorithm is available in \cite{dpskdec}.
\\If the channel phase is not available or the phase is changed as in our case by a spy, for example, the iterative decoding is assisted by a channel estimation unit (filter).
\section{Experimental Results}
\subsection{Simulation results}
In our simulation, we took an outer encoder (3,2,2) and a random interleaver. For channel phase estimation, we introduced our own values ​​directly. But, it is interesting to note that the estimation can be undertaken by a low-pass filter or a more sophisticated filter as the Wiener filter.
\subsubsection{Performance in terms of correcting power}
\begin{figure}[h]
\centering
\includegraphics[width=9cm,height=5cm]{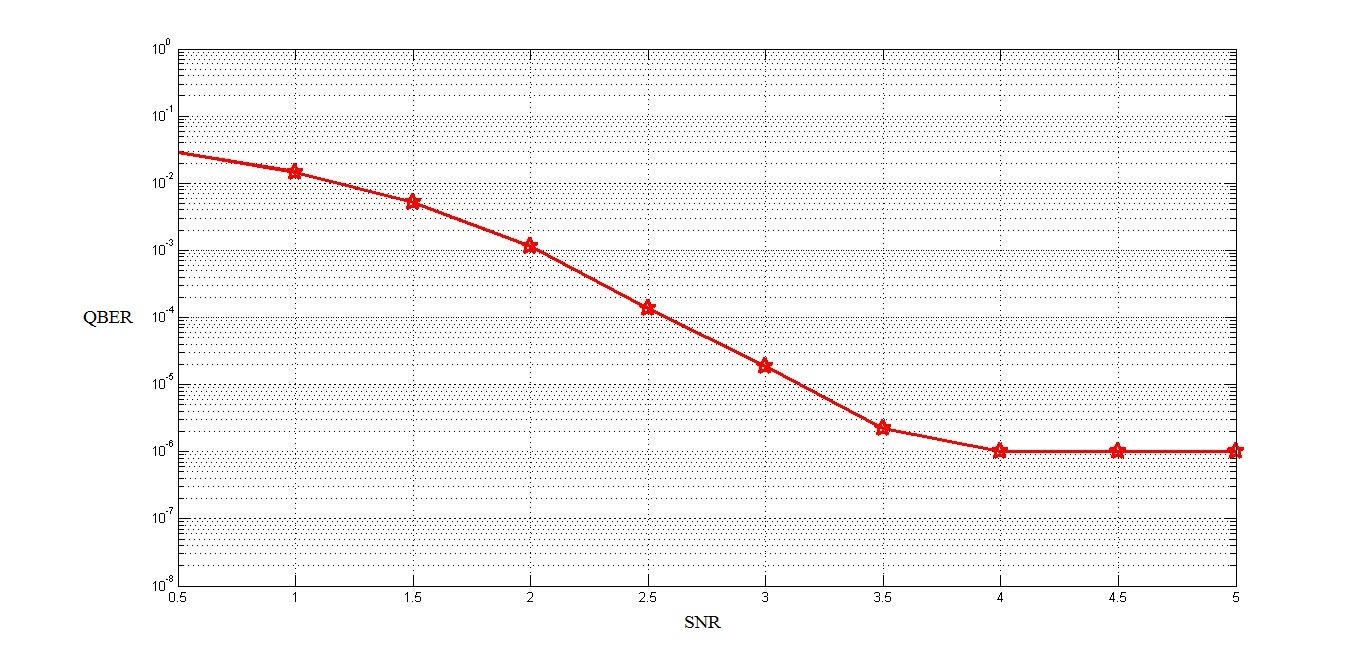}
\caption{QBER as function of $SNR$ -values after the integration of our reconciliation method.\label{Fig:c3}}
\end{figure}

It is often not easy to know exactly how the efficiency of a reconciliation method depends on the signal to noise ratio ($SNR$). However, each reconciliation technique works best for a certain range of $SNR$. In fact, after the work in \cite{preuvconti}, reconciliation SEC is recognized efficient for a $SNR$-value of about 3. But, its performance degrade at low $SNR$-value because it mainly breaks the symmetry of the gaussian problem. The Figure \ref{Fig:c3} shows the result of our simulation of the QKD protocol adapted to the case of continuous variables. Our study shows that our method can be considered suitable for a $SNR$-value starting from 1.5.

\subsubsection{Performance in terms of protocol security}
\begin{figure}[h]
\centering
\includegraphics[width=9cm,height=5cm]{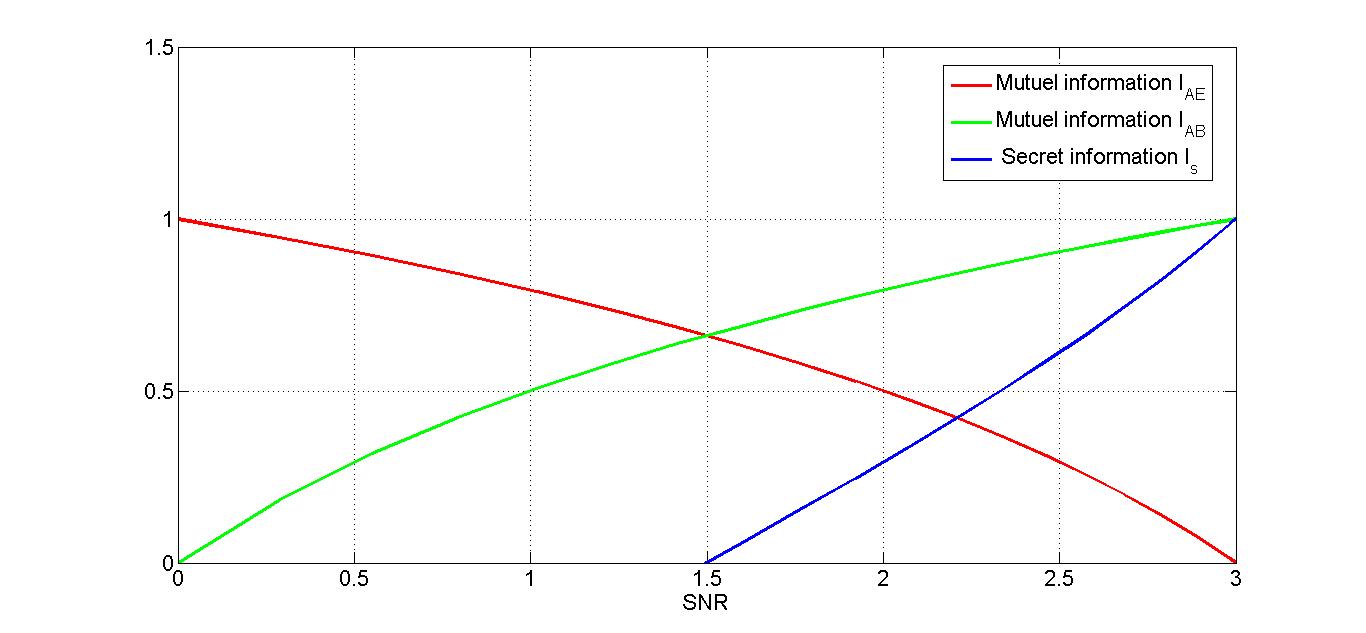}
\caption{Mutuel information as function of SNR-values.\label{Fig:info}}
\end{figure}

In our work, we use the equivalent of the protocol BB84 protocol to discrete variables. To do this, we assume that:
\begin{itemize}
\item Alice sends a series of coherent states in the quantum channel distributed with a gaussian quadrature modulation in both $X$ and $P$, variance of $V_ {A} N_ {0}$.
\item Bob randomly measures a quadrature $X$ or $P$ each received coherent state.
\item After measurement, Bob reveals publicly the quadrature he has chosen. These quadratures act as bases in the BB84 protocol to discrete variables.
\end{itemize}
In the following, we will study the security of this coherent states  protocol  by assessing the amount of secret information. We can draw the mutual information $I_{AB}$ and $I _{AE}$ and the secret information $I_ {s}$ (see Figure \ref{Fig:info}). We remark that the secret information is hardly related to the SNR. This parameter directly influence the protocol security. For example, for our simulation' settings, the QKD protocol can be considered secure at SNR starting from 1.5 as the secret information $I_ {s}$  begins to be positive.
\\Also, for the study of performance in terms of robustness against attacks, we distinguish two cases depending on the presence or absence of excess noise $\zeta$.
\begin{itemize}
  \item In the absence of excess noise, optimum attack is the beam splitter attack\cite{RC}.
  \item In the presence of excessive noise, the best attack is an entangling cloner \cite{RC}.
\end{itemize}

In both types of attacks, the spy does not change the state (amplitude, phase) but rather modifies the $SNR$. Therefore, it appears trivial if we can reduce the QBER at decreasing $SNR$, we can improve the security of the protocol. This is exactly the case of our reconciliation method.

\subsubsection{Performance in terms of complexity and computation time}
In practice, the calculation time is an important parameter for high rates and highly interactive diagrams where latency can become a problem. The integration of our method of reconciliation has a little effect on the computing time of the QKD protocol. The integration of our method adds extra time remaining of about 1.09s.  It is important to note at this stage of our study that these times are those of the execution of our scripts written in C on Intel Core i5 with 2.27 GHz processor and 4GB of RAM.  Also, the method remains low complexity since decoding is linear.  We can therefore consider the complexity as $O (n)$.
\\In summary, we can say that our method works well as far as error correcting power and contributes to the improvement the protocol security. The gain is obtained with a reasonable complexity increase.   The reconciliation method especially seems more feasible in practice. In the following, we want to show the practical side of our proposed method  by a further study on optical fiber.
\subsection{VPI Transmission Maker results}
\begin{figure}[h]
\centering
\includegraphics[width=7cm,height=4cm]{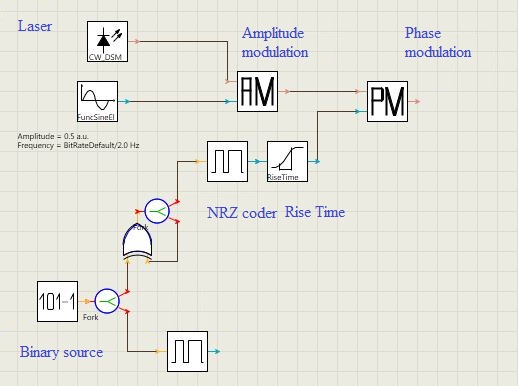}
\caption{Optical transmitter on the Alice's side.\label{Fig:alice}}
\end{figure}

\begin{figure}[h]
\centering
\includegraphics[width=7cm,height=4cm]{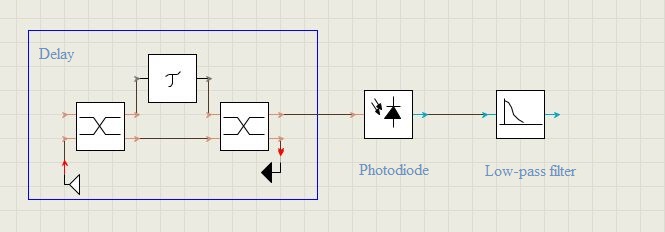}
\caption{Optical receiver on the Bob's side.\label{Fig:bob}}
\end{figure}

\begin{figure}[h]
\centering
\includegraphics[width=9cm,height=6cm]{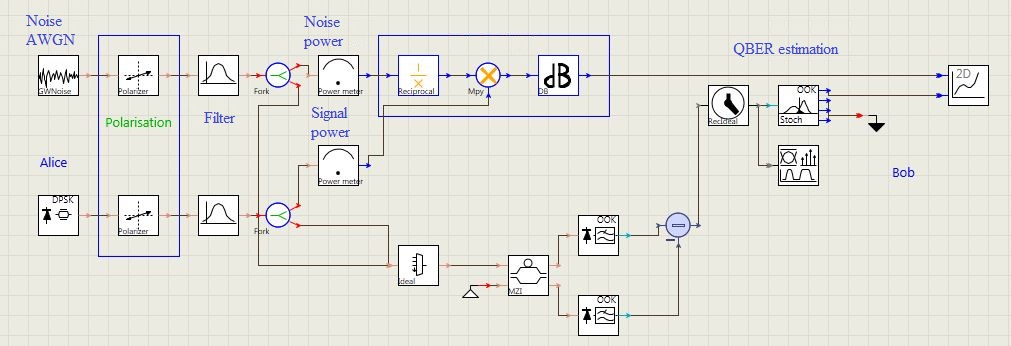}
\caption{Optical layout of a fiber optic continuous quantum cryptography system.\label{Fig:intercept5}}
\end{figure}

\begin{figure}[h]
\centering
\includegraphics[width=9cm,height=5cm]{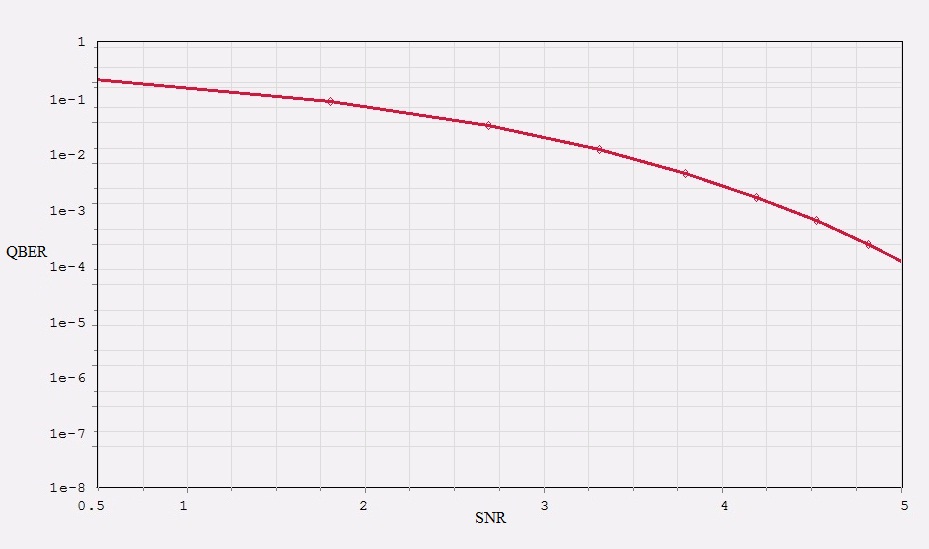}
\caption{QBER as function of $SNR$.\label{Fig:c4}}
\end{figure}

Currently, optical fiber-based technologies have already been widely deployed or at least considered to be deployed in the future access network area. The optical access networks are essentially different in the distance from the transmitter fiber on the receiver at the other end, but also in the applied modulation and multiplexing techniques and the wavelength bands used to transport signals. A QKD on optical fiber in the case of continuous variables is divided into:
\begin{itemize}
\item a DPSK modulation system for translating Alice bit into quadrature
\item a detection system which measures the arrival quadrature received by Bob
\end{itemize}
In the following, we provide descriptions for these two subsets.
The simulations to be presented thereafter, were performed using the simulation software VPI Transmission Maker.
\\Generally, in an optical link, the transmitter sends data to the transmission medium in the form of an optical signal. It is often composed of a laser diode, a modulator and a data signal generator. In a cryptographic system in the continuous case, we follow the same reasoning. As shown in the Figure \ref{Fig:alice}, we use DPSK modulation. Several studies as in \cite{DPSKproof} confirm that the DPSK modulation has a better robustness to chromatic dispersion and nonlinear effects in the optical fiber.
\\At the reception, we add a delay compensatory system of a bit of time added by DPSK modulation (see Figure \ref{Fig:bob}).
\\We analyze the influence of a noisy channel (gaussian noise) taking into account the signal to noise ratio. Thus, we consider the implementation of Figure \ref{Fig:intercept5}.
\\The Figure \ref{Fig:c4} shows the error probability without reconciliation method as a function of the $SNR$.
\\In comparison to Figure \ref{Fig:c3}, we notice that our method can reduce the quantum bit error rate. Indeed, the QBER of $10^{-3}$ is reached at $SNR = 4.23$ without our reconciliation method and $SNR = 2$ by applying our method. This gives us a gain in dB of about $2.23dB$. At QBER of $10^{-2}$, we can associate a gain of about $2dB$ ($SNR = 3$ without reconciliation and $SNR = 1$ with our method).

\begin{figure}[h]
\centering
\includegraphics[width=9cm,height=6cm]{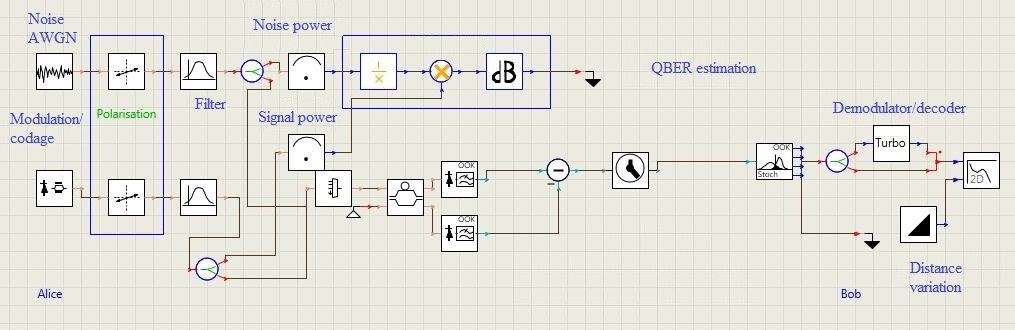}
\caption{Optical layout of a continuous quantum cryptography system with distance variance between transmitter and receiver.\label{Fig:c5}}
\end{figure}

As shown in Figure \ref{Fig:c5}, we can simulate different transmission distances by varying the attenuation.

\begin{figure}[h]
\centering
\includegraphics[width=9cm,height=5cm]{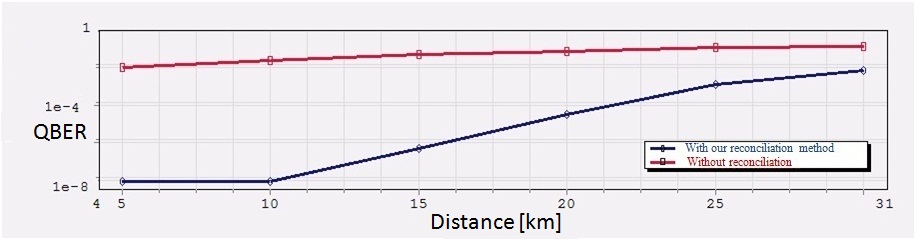}
\caption{QBER as function of distance variance between transmitter and receiver.\label{Fig:intercept6}}
\end{figure}

 We note from Figure \ref{Fig:intercept6} resulting of this simulation that the QBER increases proportionally with the length of the optical fiber.
 \\This is due to the attenuation effects of noise and of chromatic dispersion. Indeed, we can say that the attenuation in the fiber is caused by many factors. We can cite the low wavelengths which generate the intrinsic absorption of the material constituting the fiber causing increasing losses. Also, we can add other significant losses from sources such as fibers' using conditions. During the propagation, all these elements contribute to the attenuation of the light's power. Also, we can  add other elements as the chromatic dispersion which is due to the fact that light from the laser is in fact not strictly confined to the heart of the fiber. In addition, relatively high powers are supported by the optical fiber. These results in non-linear effects which will degrade the optical signal. We distinguish Kerr effects, Raman and Brillouin.
\\It should be noted that reconciliation by our method shows its limits for long distances (typically beyond 25km) because it does not consider these noises.

\section{Conclusion}
The objective of our study is to provide a cryptographic system in the continuous case through an architecture design that takes into account each parameter of the device. The system reconciles shared keys without having to discrete the quadrature assumed to be gaussian. The system can be implanted using existing devices.
 Experimental studies have proven the practical interest of the proposed reconciliation method. Unfortunately, the practical interest specially in terms of error correcting decreases proportionally with the transmission distance because of the signal attenuation.
Certainly this work has identified some research problems that can be formulated in the form of future prospects. Firstly, additional measures seems necessary to validate the approach. We can analyze other types of attack effects and other types of channels. Also, it may be wise to set the methods of reconciliation together with the surrounding system (attenuation fiber, the photodiode detection power ...).

\end{document}